\documentclass[aps,prb,10pt,twocolumn,floatfix]{revtex4}
\ifx\pdftexversion\undefined
  \usepackage[dvips]{graphicx}
\else
  \usepackage[pdftex]{graphicx}
\fi

\usepackage{amsfonts}
\usepackage{times}
\usepackage{wasysym}
\input isolatin1.sty

\begin{document}

\newcommand{\bra}[1]{\langle #1|\,}

\newcommand{\ket}[1]{|#1\rangle}

\newcommand{\braket}[2]{\ket #1\bra #2}

\newcommand{\ketbra}[2]{\langle #1| #2\rangle}

\def\dt{\,\delta t } \def\v{{\bf v}} \def\rplus{{c^+}}

\def\rminus{{c^-}} \def\chargedensity{\rho} \def\density{c}

\def\wedge{\times} 

\def\E{{\bf E}} 

\def\Eg{{\bf E}_g} 

\def\B{{\bf B}} 

\def\H{{\bf H}}

\def\Q{{\bf Q}} \def\q{{\bf q}} \def\p{{\bf p}} \def\r{{\bf r}}

\def\k{{\bf k}} \def\J{{\bf J}} \def\D{{\bf D}} \def\C{{\bf C}}

\def\O{{\cal T}} \def\r{{\bf r}} \def\kt{{\, k_BT \, }}

\def\Mesh{\Lambda}

\def\Ang{{\vec \theta}}
\def\Inertia{{I}_{\theta}}
\def\Angvec{{\bf {\bf p}_{\theta}}}
\def\Angspeed{{{\bf v}_{\theta}}}
\def\dV{{\; \rm d^3}\r}
\def\curl{{{\, \rm curl}\; }}
\def\grad{{{\, \rm grad }\; }}
\def\div{{{\, \rm div }\; }}
\def\p{{\bf p}} \def\rhat{\hat {\bf r}}

\title{
  Auxiliary field Monte-Carlo for charged particles} \author{A. C.
  Maggs} \affiliation{Laboratoire de Physico-Chimie Théorique, UMR
  CNRS-ESPCI 7083, 10 rue Vauquelin, F-75231 Paris Cedex 05, France.}
\date{\today}
\begin{abstract}
  This article describes Monte-Carlo algorithms for charged 
  systems using constrained updates for the electric field.  The
  method is generalized to treat inhomogeneous dielectric media,
  electrolytes via the Poisson-Boltzmann equation and considers the
  problem of charge and current interpolation for off lattice models.
  We emphasize the differences between this algorithm and methods based
  on the electrostatic potential, calculated from the Poisson
  equation.
\end{abstract}
\maketitle
\section {\bf Introduction}

Fast methods for calculating Coulomb interactions are of the greatest
importance in the simulation of charged condensed matter systems.
Several algorithms are available including multipole \cite{greengard}
and Fourier based \cite{darden} methods; they permit efficient
molecular dynamics simulations but are ill-adapted to Monte-Carlo
simulation. These methods have been widely used and implemented;
recent research has led to multi-time step algorithms  and aims at
generalizing the methods to multiprocessor environments. Despite this
effort the Coulomb loop is still the slowest part of many simulation
codes \cite{schlick}.

An alternative approach, the subject of this paper, is to use an
auxiliary field in order to calculate the Coulomb interaction in a
manner familiar from electromagnetism.   Many books \cite{jackson,feynman2}
 consider the following scalar functional:
\begin{equation} 
{\cal F} (\phi)=
\int \left[
{\epsilon_0 \over 2}
(\grad \phi)^2 - \rho \phi
\right]
 \dV \label{scalar} \, ,
\end{equation}
where $\rho$ is the charge density imposed by the experimental
geometry and $\phi$ is a freely variable field.  One might hope that
one could use this functional to construct alternative, and more efficient
algorithms due to the {\sl local}\/ nature of the energy function.
One is instructed to minimize Eq.~(\ref{scalar}) by taking the
functional derivative with respect to $\phi$; this leads to
Poisson's equation familiar from electrostatics:
\begin{equation}
\nabla^2 \phi_p = -\rho/\epsilon_0 \label{poisson}\, ,
\end{equation}
Thus we have a minimum principle; we
can identify the function $\phi_p$ which minimizes Eq.~(\ref{scalar}) with
the electrostatic potential.  However when we substitute the solution
 of Eq.~(\ref{poisson}) in Eq.~(\ref{scalar}) (using
appropriate boundary conditions) we find
\begin{equation} 
{\cal F} (\phi_p) = - \int {\epsilon_0 \over 2}
(\grad \phi_p)^2  \dV \,.
\end{equation}
This is the {\sl negative}\/ of the true electrostatic energy.
Recently\cite{hansen} a series of such potential functionals have been
introduced in order to treat inhomogeneous dielectric systems, such as
interfaces and pores, however all the functionals have a similar sign
defect.

This sign change is a major problem for numerical work. It implies
that the same energy can not be used to evolve the field and the
particle degrees of freedom in parallel.  Clearly if one tries to do
Monte-Carlo with $\phi$ one generates the partition function
\begin{eqnarray}
{\cal Z}(\rho) &=& \int {\cal D}\, \phi\; e^{-\beta{\cal F}(\phi)}\nonumber \\
&=& e^{-\beta {\cal F}(\phi_p)} \times {\rm const}
\end{eqnarray}
as can be seen by substituting $\phi=\phi_p+\tilde \phi$. This is the wrong
statistical weight for particles interacting via Coulomb's law. While
evaluation of the energy {\sl seems}\/ to have been reduced to a local
calculation in practice a {\sl global}\/ optimization is needed every
time step to minimize Eq.~(\ref{scalar}) and then flip the sign of the
result.  For Monte-Carlo this leads to such a level of inefficiency
that the method is useless.

This sign problem is most surprising: We are aware that Maxwell's
equations allow the co-evolution of field and particles with the
correct sign of the electromagnetic energy.  What has gone wrong in
the formulation of electrostatics in terms of an auxiliary field? The
solution to the paradox is found by recognizing that the correct sign
of interaction is only possible with a {\sl vector}\/ field, a point
made forcefully in Ref. \cite{feynman} in a general discussion of
the forces of nature.

\section{Vector Functionals}
Recently\cite{acm} an efficient Monte-Carlo algorithm for charged
systems was formulated using electric field lines (rather than the
electrostatic potential) as the true dynamic degree of freedom.  The
algorithm consists of a local energy functional together with {\sl
  local}\/ update rules; no global optimization is needed.  Coulomb
interactions result from the energy
\begin{equation}
{\cal U} = {\epsilon_0 \over 2} \int \E^2   \dV
\label{Energy} \, ,
\end{equation}
where the electric field, $\E$ is constrained by Gauss' law,
\begin{math}
  \div \E -\rho/\epsilon_0=0
\label{constraint}\,.
\end{math}
Introduction of a Lagrange multiplier for the constraint implies that
the minimum energy is indeed
\begin{equation} 
{\cal U}_p (\phi_p) =  \int {\epsilon_0 \over 2}
(\grad \phi_p)^2  \dV \, ,
\end{equation}
with $\phi_p$ (now the Lagrange multiplier) again the solution to
Poisson's equation.

We now show that the use of the energy Eq.~(\ref{Energy}) at finite temperature still
leads to pure Coulomb interactions.
 Let us define a partial partition function by integrating over the
electric field but {\sl not}\/ the particle positions, ${\bf r}_i$:
\begin{equation}
{\cal Z}(\r_i) =  \int {\cal D} \E\; \prod_{\bf r} \delta(\div \E -\rho({\bf r}_i)/\epsilon_0)\;
e^{-\beta {\cal U} } \label{Z} \,.
\end{equation}
Let us evaluate the integral by changing variables, $ {\E}_{tr} = \E
+\grad \phi_p$. We find that
\begin{eqnarray}
{\cal Z}(\r_i) &=& \int {\cal D}{\E}_{tr}\;  \prod_{\bf r}\; \delta({\div  \E}_{tr}) 
e^{
-\beta {\epsilon_0\over 2} \int (
{\E}_{tr} - \grad \phi_p
)^2 \dV
  }\nonumber \\ 
&=& e^{-\beta {\epsilon_0 \over 2} 
\int (\grad \phi_p )^2} \times\\
&\int& {\cal D}{\E}_{tr}\;   \prod_{\bf r}  \delta({\div  \E}_{tr}) 
e^{
-\beta {\epsilon_0\over 2} \int
{\E}_{tr}^2 \dV} \nonumber \\
&=& {\cal Z}_{Coulomb}({\r_i}) \times {\rm const} \nonumber
\end{eqnarray}
The cross term in the energy, $\int {\E}_{tr}\cdot \grad \phi_p \dV$, is
identically zero as can be seen by integration by parts.  We discover
that integration over the tranverse modes multiplies the statistical
weight of each configuration by a constant.  The longitudinal
component of the electric field, $-\grad \phi_p$, produces an
effective Coulomb interaction between the particles.

We shall now give a full description of the algorithm showing how
configurations can be dynamically generated according to the
constrained statistical weight in Eq.~({\ref{Z}}). This gives rise to
a lattice gas in which charges are confined to a grid. We demonstrate
numerically the efficiency of the algorithm for a system of lattice
dipoles.  A coarse graining of the electric degrees of freedom leads
to a new algorithm for treating the Poisson-Boltzmann equation as well
as a {\sl non-constrained}\/ algorithm for the Yukawa potential.  We
then show how to treat off lattice models by interpolating charges to
the grid.  We show that the generalization of the algorithm to
inhomogeneous dielectric media automatically re-sums a large class of
fluctuation based potentials such as the Keesom potential, capturing
important characteristics of polar  dielectric materials which are
missed by the Poisson equation.

\section{Periodic Boundary Conditions}
\subsubsection{Conventional approach}
Imposition of periodic boundary conditions in systems interacting with
Coulomb interactions is subtle due to the conditional convergence of
Coulomb sums \cite{leeuw1}: When extrapolating periodic images of the
system to infinity there is a residual dependency on the dielectric
constant, $\epsilon_s$ of the surrounding medium.

The total potential, $\Phi$ is the sum of $\phi_p$, coming from the
solution of the Poisson equation with periodic boundary conditions, and
a dipolar contribution: $\Phi= \phi_p + \phi_{d}$.  Here,
\begin{equation}
\phi_{ d} = - \bar \E \cdot \r \, ,
\label{phid}
\end{equation}
where $\bar \E$ is a vector proportional to the dipole moment per unit
volume of the simulation cell, ${\bf d}_E$ \cite{leeuw1}.
\begin{equation}
\bar \E =  -{1\over \epsilon_0 (1 + 2 \epsilon_s)} {\bf d}_E
\label{ebardef}
\end{equation}
This constant field reminds us of the ``depolarizing field'' in
elementary treatments of spherical dielectric media\cite{jackson}.
Two common choices of boundary condition are ``tin-foil'' with
$\epsilon_s= \infty$ and ``vacuum'', $\epsilon_s=1$. 
Clearly the tin-foil limit corresponds to keeping just the periodic
potential, $\phi_p$.

The field $\bar \E$ generates a purely additive contribution to the
energy density:
\begin{equation}
{ u}_0 =  {{\bf d}_{E}^2  \over 2 \epsilon_0 (1 + 2 \epsilon_s)}
\end{equation}
This extra energy term can be awkward to handle in molecular dynamics;
its definition leads to discontinuities in the energy function since
${\bf d}_E={1\over V}\sum_i e_i \r_i$ is calculated with a given,
fixed Bravais lattice; $V$ is the simulation volume, $e_i$ the
charge of particle $i$ at position $\r_i$ in the simulation cell.

\subsubsection{Constrained algorithm}
The general solution to Gauss' law in periodic systems is
\begin{equation}
\E= -\grad \phi_p + \curl \Q +{\bar \E}
\end{equation}
where $\E_{tr}=\curl \Q$ is an arbitrary transverse field.  Our algorithm is
closely related to the following Maxwell equation:\cite{acm2}
\begin{equation}
\epsilon_0 {\partial \E_{\rm} \over \partial  t} 
= -\J + \curl \H \label{maxwell} \, ,
\label{ampere}
\end{equation}
with $\J$ the electric current and $ \H$ the magnetic strength. If we
start a simulation in a state in which $\bar {\E}=0$ then an
evolution of the field according to Eq.~(\ref{maxwell}) (as well, it
will be seen, as with our algorithm) generates fields with
\begin{equation}
\bar \E(t) = - {1 \over V}\int {\rm d}t\, \int \dV\; \J/\epsilon_0 =
{\bf -d}/\epsilon_0
\label{ebar}
\end{equation}
so defining ${\bf d}$.  We have used
the fact that the integral of the $\curl$ of a periodic function is
zero.  This is {\sl very similar}\/ to the Ewald convention of Eq.
(\ref{ebardef}) with $\epsilon_s=0$. We note that this choice is
rather unconventional, one normally considers $1< \epsilon_s< \infty$.

If the charges in the simulation are bound (so as to remain in the
basic unit cell) then ${\bf d} ={\bf d}_E$.  If there are free charges
in the simulation, $\pm e$, then the same configuration can be
produced by different currents which wind about the periodic system:
We only require that ${\bf d} -{\bf d}_E=e{\bf b}/V$, where ${\bf b}$
is a Bravais lattice vector.  We again find that there is an additive
contribution to the energy density of the form
\begin{equation}
u_0 = {  {\bf d}^2 \over 2\epsilon_0}
\end{equation}
We note the natural use of ${\bf d}$ rather than ${\bf d}_E$ is a
great advantage for molecular dynamic simulation. It leads to energies
which are continuous functions of the particle position.

In our simulations of periodic systems 
we are also capable of reproducing tin-foil boundary
conditions. We do this by adding a constant electric field
$\E_g$ to the zero wavevector component of the electric field.
so that $\bar \E = -{\bf d}/\epsilon_0 +  \E_g$. This leads to a simple
modification of $u_0$,
\begin{equation}
u_0 = {  {({\bf d}-  \epsilon_0 \E_g)}^2 \over 2\epsilon_0}
\end{equation}
Integration over $\E_g$ as well as the transverse field now leads to
an effective statistical weight which varies as
\begin{math}
  e^{-\beta/2 \int (\grad \phi_p )^2 \dV}
\end{math}.
The particles interact through the periodic solution to Poisson's
equation, $\phi_p$.

Finally in this section we note that a consistent description of
polarization effects in quantum mechanics (based on considerations of
gauge invarience) require the use of $\bf d$ rather than
${\bf d}_E$ for the polarization \cite{vanderbilt}.

\section{A scalar functional via gauge transformations}

In this section we continue the  development of our
formulation of the electrostatic potential in order to show that despite the
failure of the variational principle Eq.~(\ref{scalar}) it is possible
to link the electrostatic energy Eq.~(\ref{Energy}) to a scalar
potential.  This section is independent of the practical implementation
of the algorithm and may be omitted on first reading.

Our separation of the electric field into longitudinal and transverse
components is similar in many ways to the Coulomb gauge in
conventional electromagnetism.  It leads, however, to a description of
the dynamics in terms of potentials which is non-local; as in
conventional electromagnetism other choices of gauge are possible.

Let us write the electric field in an arbitrary gauge in the form
\begin{equation}
\E = - \grad \phi -{\bf A}
\end{equation}
From Gauss' law we deduce that
\begin{math}
  \rho/\epsilon_0 = - \nabla^2 \phi - \div {\bf A}
\end{math}. We now apply a gauge condition
\begin{equation}
\div {\bf A} + \alpha {\partial \phi \over \partial t } =0
\label{gauge}
\end{equation}
for some constant $\alpha$.  We discover that the propagation of the
potential $\phi$ is now local:
\begin{equation}
{\alpha} {\partial \phi \over \partial t} = \nabla^2 \phi + \rho/\epsilon_0 
\end{equation}
The dynamics of the electric field in the algorithm are related to the
Langevin equation \cite{acm2}
\begin{equation}
  {\partial \E \over \partial t} = 
 \left (\epsilon_0 \nabla^2 \E - \grad \chargedensity \right)/\xi  -\J/\epsilon_0 +\vec \zeta(t)
\label{basic}
\end{equation}
where $\xi$ sets the time scale of the dynamics and $ \vec \zeta(t)$
is a transverse noise.  From these equations we see that the special
choice $\epsilon_0 \alpha= \xi$ allows a particularly simple equation
for the vector potential $\bf A$.
\begin{equation}
 {\partial {\bf A} \over \partial t} = {1\over \alpha} 
\nabla^2 {\bf A} + {\bf J}/\epsilon_0 +\vec \zeta(t)
\end{equation}
In this diffusive gauge we calculate the electric energy
Eq.~(\ref{Energy}) in terms of the potentials
\begin{equation}
{\cal U} =  \int \left\{ \rho \phi - {\epsilon_0 \over 2 }(\grad \phi)^2 +
 {\epsilon_0 {\bf A}^2 \over 2} \right\}
 \dV
\end{equation}
Cross terms in ${\bf A}\cdot \grad \phi$ have been eliminated with
help of the gauge condition and by using the equation of motion for
the scalar potential. We recognize that this functional is closely
related to the naive scalar functional Eq.~(\ref{scalar}); it is,
however, dynamically constrained by Eq.~(\ref{gauge})

\section{Algorithm}

The Monte-Carlo algorithm uses the Metropolis criterion together with
the constrained energy function Eq.~(\ref{Energy}). In order to
generate configurations according to the statistical weight of
Eq.~(\ref{Z}) we need to
\begin{itemize}
\item move particles without violating the constraint of Gauss' law to
  preserve the delta function on $\div \E-\rho/\epsilon_0$.
\item integrate over the transverse degrees of freedom ${\E}_{tr}$, of
  the electric field
\item integrate over $ \E_g$.
\end{itemize}
Our third Monte-Carlo step eliminates the slaving of $\bar \E$ to the
current.  If we perform simulations in which we do not integrate over
the variable $\E_g$ we find a summation of the Coulomb energy which
is {\sl different from}\/ that generally used in the Ewald method, but
identical to that implied when integrating the Maxwell equations.

\subsubsection{Implementation}

The system is discretized by placing charged particles on the $N=L^3$
vertices of a periodic cubic lattice, $\{i\}$. The components of the
electric field $E_{i,j}$ are associated with the $3N$ links $\{i,j\}$
of the lattice. There are $3 N$ plaquettes on the lattice each defined
by four links. We use the notation $E_{1, 2}$ to denote a local
contribution to electric flux leaving $1$ towards $2$. If the link
from $i$ to $j$ is in the positive $x$ direction we consider that this
is the local value of the $x$ component of the field $\E$. The $3N$
variables $E_{i,j}$ are thus grouped into $N$ three dimensional
vectors.

It is convenient to consider Gauss' law in the equivalent integral
form
 \begin{equation}
   \int \E \cdot\, d {\bf S} = e_i/\epsilon_0
\label{integconstraint}\,,
\end{equation} 
where $e_i$ is the total charge at the site $i$, enclosed by the
surface integral, which we discretize as
\begin{equation}
 \Mesh^2 \sum_{j} E_{i, j} = e_i/\epsilon_0 \label{sum} \,.
\label{discgauss}
\end{equation}
$\Mesh$ is the lattice spacing.  The sum in Eq.~(\ref{sum})
corresponds to the total electric flux leaving the site $i$ through
the surrounding plaquettes. Since we are considering $E_{i,j}$
as a directed flux we have $E_{i,j}= -E_{j,i}$.

In order to integrate over $\bar{\E}$ we add a global background
vector $\Eg$ to $E_{i,j}$. The background field does not contribute to
the divergence of the electric field, it does, however contribute to
the energy
\begin{equation}
{\cal U}={\Mesh^3 \epsilon_0\over 2}\sum_{links} 
(E_{i, j}  +  E_{g}^\alpha)  ^2 \label{shiftenergy}\,,
\end{equation}
with $E_{g}^\alpha$ the component of the background field parallel to
the link $i,j$.  For convenience we also store the sum of the
variables $E_{i,j}$, we denote this single 3-vector $\E_{total}$.  For
instance the $x$ component, $E_{total}^{x} = \sum_{\rm{x }} E_{i,j}$,
where the sum is over all $x$ directed links. We note that $\bar
\E=\E_{total}/N+\E_g$.

With this discretization of the electric field we check the derivation
of the Coulomb interaction between charged particles. Particles now
interact via a lattice Green function, rather than with the continuum
Green function of $G_c=1/r$: The interaction is identical to that found
from solving the Poisson equation with a simple difference scheme on
the same lattice. 
Explicitly we find that
\begin{equation}
G_l = \int  {{\rm d^3} {\bf q} \over (2 \pi)^3} \; {e^{i {\bf q}\cdot {\bf r}}
\over 6- 2 \cos {q_x} -2 \cos{q_y} -2 \cos{q_z}}
\end{equation}
 The lattice also regularizes the self energy of the
particles which scales as $e^2_i/\epsilon\Mesh$.  This self energy
is analogous to the Born self energy important in solvation theory.

\subsubsection{Particle Motion}
The simulation starts with Gauss' law satisfied as an initial
condition.  We displace a charge, $e$ situated on the lattice site,
$1$, to the neighboring site, $2$; for figures and more detail we
refer the reader to our previous work\cite{acm}. The constraint is
again satisfied after motion of the particle if the field associated
with the connecting link is updated according to the rule $E_{1,2}
\rightarrow E_{1,2} - e/\epsilon_0\Mesh^2$.  The trial is accepted
or rejected according to the Metropolis algorithm. If the move is
accepted we must remember to modify $\E_{total}$. It is this update
that leads to the slaving of $\bar {\E}$ to the current,
Eq.~(\ref{maxwell}), since it corresponds to $\epsilon_0 \dot \E =-\J
$

\subsubsection{Plaquettes}
We integrate over all transverse modes in the partition Eq.~(\ref{Z})
by modifying together the four link field values of one of the $3N$
plaquettes.  In such an update the sum of the entering and leaving
fluxes at each site does not change.  This update modifies the
electric field by a pure circulation; $\E_{total}$ does not change. We
perform this update by choosing one of the $3L^3$ plaquettes randomly.
The amplitude of the trial update is uniformly distributed between
$-\theta_0$ and $\theta_0$ where $\theta_0$ is chosen to have an
acceptance rate of between 40\% and 60\%.

Let us note that the distinction between link and plaquette degrees of
freedom is very similar to that found in the Yee algorithm for
integrating Maxwell's equations \cite{yee}.
\subsubsection{Background field}

The Metropolis algorithm for the variable $\E_g$ is assured by using
the energy Eq.~(\ref{shiftenergy}). However, since the individual link
fields, $E_{ij}$, are unchanged when we modify $\E_g$ we calculate the
{\sl change}\/ in energy using the simplified expression
\begin{equation}
{\cal U}_g= {\Mesh^3 \epsilon_0 \over 2} 
(N   \E_g ^2 + 2  \E_g \cdot \E_{total}) \,.
\end{equation}
By following the evolution of $\E_{total}$ we avoid having to
calculate the sum of the electric fields; modification of the global
field is possible in a time which is $O(1)$.

\section{Efficiency}

In \cite{acm2} we gave an demonstration of the efficiency of the
algorithm for a charged lattice gas; in this section we hope to
convince the reader that the algorithm is equally efficient in
simulating dipolar systems in the absence of Debye screening.

To model a dipolar system we take a lattice gas of charged (and
self-avoiding) particles as described above and pair each positive
charge with a negative charge: They are linked by a harmonic, zero
length, spring so that the elastic energy between them is given by
${\cal U}_{spring} = \gamma ({\bf r}_i-{\bf r}_j)^2/2$, ${\bf r}_i$ is
the position of the particle $i$ on the lattice, $\gamma$ the spring
stiffness. During the simulation we record the Fourier components of
several important quantities: the charge density, the particle density
and the transverse electric field (that part of the field for which
the discrete divergence Eq.~(\ref{discgauss}) is zero). From these
recordings we deduce the two-time correlation functions.  We find that
the temporal correlations of the Fourier components are well described
by single exponentials, as described previously \cite{acm2}.

\begin{figure}[htb!]
  \includegraphics[scale=.45] {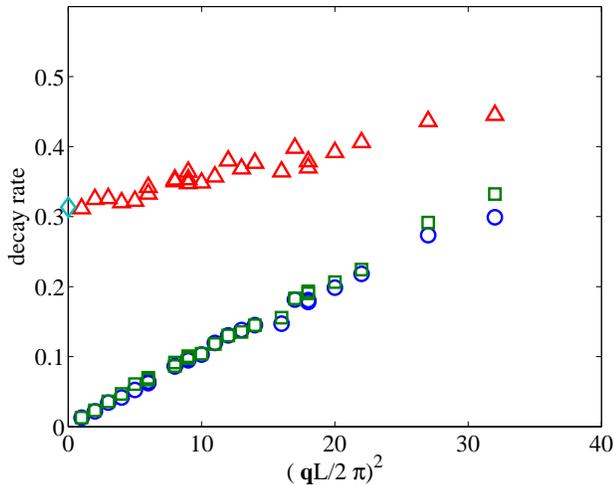}
  \caption{
 Decay rate, in particle-sweeps, 
  of correlation functions as a function of wavevector.  $\square$:
  density fluctuations, $\triangle$: charge fluctuations, $\octagon$:
  transverse electric field.  60,000 particle-sweeps, 8 recordings per
  sweep.  $\Mesh=1$, temperature $T=.5$, $|e|=1$, $\gamma=1/2$,
  $\epsilon=1$. The single mode at ${\bf q}=0$ describes the dynamics
  of $\bar { \E}$.
}
\label{figa}
\end{figure}

We now plot the decay rate of each mode measured in
``particle-sweeps'' as a function of wavevector, 
Fig.~(\ref{figa}).
One particle-sweep is defined as the time needed to attempt (on
average) one update for each {\sl particle} in the simulation,
independent of the number of Monte-Carlo trials on other degrees of
freedom; we update the plaquettes with various rates in order to study
the effect on the dynamics.  Measuring time in this way allows us to
study the intrinsic dynamics of the dimers and their coupling to the
transverse field. Note that the computational effort needed to move a
particle or to update a plaquette is almost the same; both are
dominated by the time needed to generate random numbers and calculate
the exponential function needed for the Metropolis algorithm. In our
simulations we used a lattice of $L^3=20^3$ sites and 24,000
plaquettes with 1200 dimers (2400 particles).

In looking at 
Fig.~(\ref{figa})
 one should first note that there are two diffusive modes,
with the decay rate proportional to ${\bf q}^2$, where ${\bf q}$ is
the wavevector of the mode.  They correspond to density fluctuations and to
electric field diffusion, described by Eq.~(\ref{basic}).  The third
mode, describing charge fluctuations, has a finite frequency at long
wavelengths; it couples to the internal modes of the dimers which
relax rapidly due to the spring.

For illustrative purposes we have tuned the relative probabilities of
particle motion and plaquette updates in 
Fig.~(\ref{figa})
 so that the
diffusion rates of the density and the transverse electric field are
very similar.  To do this we used a relative probability (per degree
of freedom) of particle motion to plaquette updates of $1: 1/3$; Since
there are considerably more plaquettes than particles about 75\% of
the time in the simulation is devoted to updating the transverse
field. For simplicity we do not update $ { \E}_g$;
we are simulating with ``Maxwell'' boundary conditions.

\begin{figure}[htb!]
  \includegraphics[scale=.45] {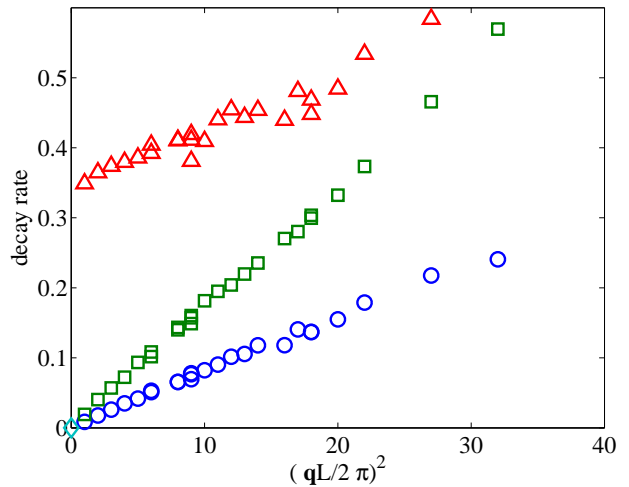}
\caption{  As Fig.~(\ref{figa}) but with charge-free dimers. Charge
  correlation function replaced by partial structure factor sensitive
  to internal dimer motion. The dimers are diffusing faster.}
\label{figb}
\end{figure}

In
 Fig.~(\ref{figb})
 we simulate a similar system with the same ratio
of particle motion to plaquette updates. The only modification is that
the charges on the dimers are now put to zero; there is no charge in
the simulation. We note two points in comparison with
Fig.~(\ref{figa}):
 The diffusion rate of the dimers has increased by
some 60\%, while the diffusion rate of the electric field has slightly
fallen. In this plot we replace the charge-charge correlation function
by an analogous partial structure factor, in which the scattering
amplitude of one particle in a dimer is $+1$, while the second
particle has a scattering amplitude of $-1$.  This structure factor is
again sensitive to the internal dynamics of the dimer.

\begin{figure}[htb!]
  \includegraphics[scale=.45] {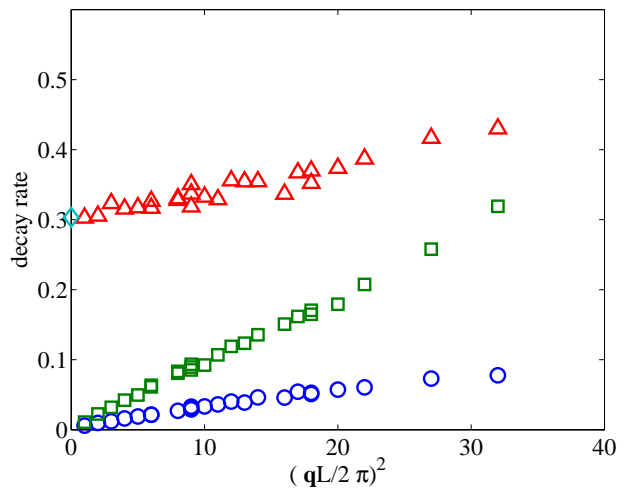}
\caption{ As Fig.~(\ref{figa}) but with fewer updates to transverse modes.
  Dimers carrying charges.  The relaxation dynamics of density
  fluctuations and the charge fluctuations are only slightly modified
  despite the order of magnitude decrease in the work done on the
  plaquettes.  }
\label{figc}
\end{figure}

In 
Fig.~(\ref{figc})
 we again simulate charged dimers, we reduce
however the number of plaquette updates in the simulation so that they
now occur in the relation $1: 1/30$. In this simulation 75\% of the
time is spent updating the position of the particles, only 25\% of CPU
time is devoted to the transverse field. As might be expected the
propagation of the transverse modes has slowed down. However the
equilibration of the charge and density modes is not greatly modified
in comparison with 
Fig.~(\ref{figa}).

\begin{figure}[htb!]
  \includegraphics[scale=.45] {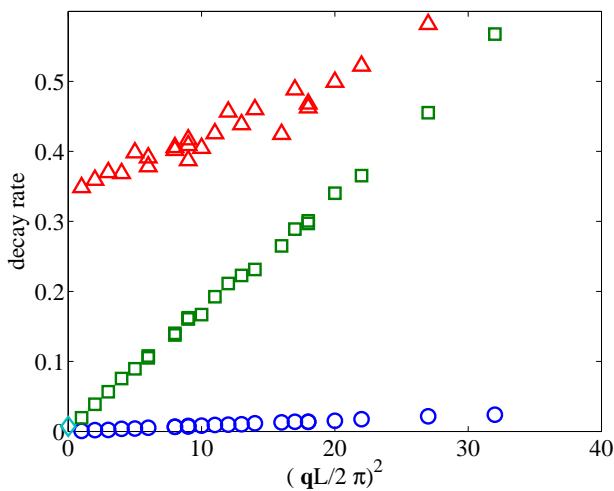}
\caption{ As Fig.~(\ref{figa}) but with fewer updates to transverse modes
  and charge-free dimers. The dynamics of the transverse modes have
  slowed in comparison with Fig.~(\ref{figc}) in the absence of
  coupling to the charges.}
\label{figd}
\end{figure}

In 
Fig.~(\ref{figd})
 we have simulated charge-free dimers with
plaquettes updated in the proportion of $1: 1/30$. We note that
compared with 
Fig.~(\ref{figc}) 
(where the dimers carry charges) the
transverse modes have substantially slowed. Motion of the particles is
playing an important role in agitating the field variables in
Fig.~(\ref{figc});
 this accelerates the propagation of the electric
field.

From our results on the simulation on this simple dimer system we find
that quite remarkably one can ``get away'' with very little work on
the plaquette degrees of freedom. How can this be so?  Motion of a
particle along the four links defining a plaquette is equivalent to a
single plaquette update; the motion of the particles is, on its own,
quite good at integrating over the transverse degrees of freedom.
This integration is clearly imperfect
since the only values of the plaquette circulation that
are generated are integral multiples of $e / \epsilon  \Mesh$.
We see, however, that the plaquette updates only have to
``fill in'' between the integral multiples of $e/\epsilon\Mesh$
rather than perform the full integration of these modes. 

It is also
rather natural that longitudinal and transverse modes (with scalar and
vector symmetries) are {\sl weakly coupled} in a homogeneous system at
long wavelengths; the lowest order couplings occurring in the free
energy are presumably at least cubic; if the number density
fluctuation is $n$ and the transverse field fluctuation ${\bf e}_{tr}$
one could expect that the lowest order coupling in a Landau picture is
${\bf e}_{tr}^2 n$.

Examining the figures we see that the total overhead in simulating the
charged dipole, compared to a charge-free dimer is about a factor 2
with appropriately chosen simulation parameters.

The simulation of 
Fig.~(\ref{figa})
 corresponds to 30 minutes of
simulation on a pentium 4 2.5Ghz.  Total clock time was (much) longer
due to the overhead of the Fourier analysis of the mode structure.

We note that the temperatures used in these simulations is {\sl very} high
compared with those of interest in condensed matter physics.  The
Bjerrum length is comparable to the particle size in these simulations
at a temperature of about $T=1/4 \pi$, the present simulations are
however at $T=1/2$.  We find that at low temperatures the mobility of
the particles drops very strongly, since there is a finite barrier for
particle motion due to the discrete nature of the charges and the
lattice.  We have recently implemented an off lattice
version of the algorithm and find that we can easily overcome this
artefact of the discretization \cite{joerg}.

\section{Poisson-Boltzmann simulations}

Classic methods for treating the Poisson-Boltzmann equation require a
calculation of the long ranged Coulomb interaction, either in real
space or in Fourier space \cite{lowen}. In this section we show how
constrained Monte-Carlo gives rise to a local treatment of the
Poisson-Boltzmann equation. The method, however, has some original
features since rather than considering the minimum of the
Poisson-Boltzmann functional we integrate over the charge degrees of
freedom, creating a fluctuation enhanced Poisson-Boltzmann
algorithm\cite{netz}.

\subsection{Formulation of the free energy}

Consider a system of $n$ particles in a fixed volume $V$ with
coordinates $\q$.  The partition function is
\begin{equation}
{\cal Z} = \int e^{-\beta \cal U} \; d\q\,,
\end{equation}
where $\cal U$ is the configurational contribution to the energy and
the integral is over all particle positions.  Let us coarse grain the
evaluation of the partition function by subdividing the volume $V$ in
to a large number of elementary cells of volume $\Mesh^3$, replacing
the configuration integrals by sums. In each cell let there be $n_i$
particles. Then
\begin{equation}
{\cal Z} = \sum n! \prod_i 
\left( {\Mesh^{3n_i}  \over n_i!} \right) e^{-\beta \bar{\cal U}}\,.
\end{equation} 
Here the combinatorial factor counts the number of ways of
distributing particles between boxes and the energy has been
re-expressed in a coarse grained manner, ${\cal U}=\bar {\cal U}(n_i)
$, in terms of the cell occupation numbers.  The factor
$\Mesh^{3n_i}$ comes from the $n_i$ fold integral over the cell
volume.  Re-expressing the $n_i$ in terms of local concentrations of
particles, $\rho_i = n_i/\Mesh^3$, and exponentiating the
combinatorial factor (using $\log(n!) \approx n\log n-n$ ) we find the
effective energy function for the coarse grained description
\begin{equation}
{\cal H} = \kt \Mesh^3 \sum_i 
(\rho_i  \ln (\rho_i \Mesh^3) -\rho_i) +\bar{\cal U}(\{\rho_i\})\,,
\end{equation}
where the sum over $i$ is now a sum over both boxes and chemical
species.  This expression is very similar to that used in many mean
field studies, but the derivation shows that this effective energy
function is valid for {\sl fluctuating}\/ densities.  It is {\sl
  not}\/ a functional simply valid as a minimum principle as is
emphasized in density functional approaches to the Poisson-Boltzmann
equation.  In our simulations the variables $\rho_i$ are fluctuating
variables.

In order to describe a charged coarse grained fluid we must choose
$\Mesh$: The ions are separated on average by a distance $\xi=
\rho^{-1/3}$ while the Bjerrum length is given by $\ell_b = e^2/4 \pi
\kt \epsilon$. 
We also introduce a further length scale: the Debye length, the length
scale at which the energy stored in the electric field, due to
concentration fluctuations, is equal to $\kt$; we find that $\ell_d^2
\sim \xi^3/\ell_b$.  In order to use the coarse grained entropy we
require that $\Mesh \gg \xi$ so that $n_i\gg1$, however if we wish
to describe screening in detail we also require $\Mesh \ll \ell_d$
implying a relatively narrow window of possible values for the coarse
graining parameter.

\begin{figure}[htb!]
  \includegraphics[scale=.45] {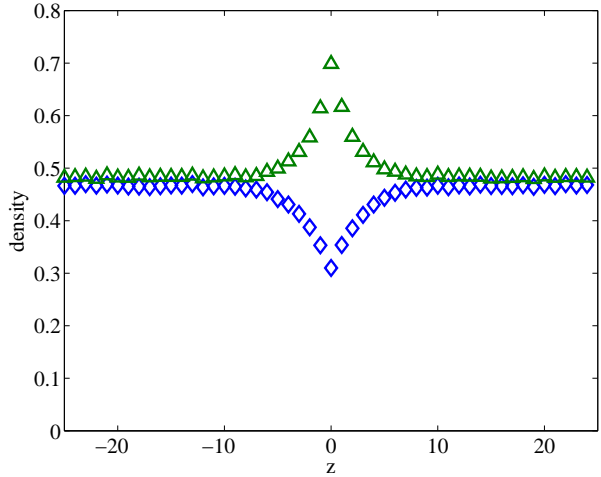}
\caption{A positively charged plane is simulated in in the presence of counter-ions
  using the Poisson-Boltzmann formulation eq(\ref{pb}). We plot the
  concentration of the of two, symmetric ion concentrations as a
  function of the distance from the plane. System of size
  $50\times50\times50$. Concentration is for a single equilibrated
  system. $\triangle$ negative charges, $\lozenge$ positive charges.}
\label{poisson1}
\end{figure}

The full system we simulate is
\begin{eqnarray}
{\cal H} &=& \kt \Mesh^3 \sum_{i} ( \rho_i \ln \rho_i \Mesh^3 -\rho_i) \label{pb}\\
 &+&
 {\Mesh^3 \over 2} \epsilon_0 \sum_{links}{E_{i,j}^2 } \, , \nonumber
\end{eqnarray}
where the electric field is constrained by Gauss' law, with both
external charges and $\rho_i$ acting as sources.

Other contributions to the free energy varying in $\rho^{3/2}$ are
often introduced as local field corrections in density functional
theories.  However, such energies are derived on a length scale larger
than the Debye length and are due to correlations of the electric
field over $\ell_d$. In the effective free energy measured on a scale
$\Mesh< \ell_d$ such terms should not be added by hand.  They will
be generated automatically by the algorithm via fluctuations in the
charges.

The derivation of the effective Hamiltonian on the scale $\Mesh$ is
rather close in spirit to the renormalization group. It would be
interesting to perform the numerical renormalization group on a
charged fluid in order explicitly measure the effective Hamiltonian as
a function of the scale for $\xi<\Mesh<\ell_d$. The derivation
breaks down in some particularly interesting cases, including strongly
charged surfaces. In this case a new, independent, length, the
Gouy-Chapman length becomes smaller than the coarse graining cell.

As a test of the algorithm we simulated,
 Fig. (\ref{poisson1}),
 a uniform positively charged plane in a background of positive and
 negative ions.  We see the depletion of the positive ions and the
 attraction of negative ions near the surface.  In the simulation the
 plane was stationary however the method allows parallel dynamics of
 both the ionic and the macro-ion degrees of freedom without having to
 perform global updates or optimizations.  Such simulations are useful
 in the simulation of flexible charged objects such as
 polyelectrolytes within the Poisson-Boltzmann approximation; in such
 flexible objects we can individually update the position of
 individual charges in the macro-ion as described in the application
 to electrolytes.  The method is less likely to be useful in the
 simulation of large rigid objects such as a colloidal sphere
 carrying large charges. In such cases the simultaneous modification
 of $\E$ on many links leads to a low Monte-Carlo acceptance rate, or
 when working off-lattice small step sizes.  Our method is also
 incompatible with smart global algorithms, such as the pivot
 algorithm in polymer simulation, for similar reasons.

In the functional Eq.~(\ref{pb}) we integrate over the continuous
variables $\rho_i$. However, numerically it is just as easy to keep
the discrete charges, $n_i$. In such a way one still describes a
coarse grained system so that $\langle n_i \rangle >1$ but a higher
degree of detail is retained. A improved account for the entropy is
also possible if a better approximation for $\log{(n!)}$ is used. It
would be particularly interesting to understand if attractive,
correlation effects can be reproduced within such a coarse grained,
discrete model.

Finally we wish to make some remarks about the ergodicity of the
Poisson-Boltzmann algorithm: We have already seen in our dipolar model
that the motion of the charges excites transverse degrees of freedom
of the field. It was only the discreet nature of the charges in the
dipolar system which prevented the transverse field from coming to
equilibrium even in the absence of independent field updates. In the
Poisson-Boltzmann algorithm the amount of charge transfered at each
step is randomly chosen: the algorithm is ergodic {\sl even without
  the field updates} which can be dropped from the algorithm.

\subsection{Screened Coulomb Interactions}

In even coarser-grained treatments of charged systems the counter ion
degrees of freedom are eliminated entirely by using an effective
Yukawa interaction $e_1 e_2\exp(- \kappa r)/r$.  One  implements this
potential numerically by introducing a {\sl non-constrained}\/ energy
function for the electric field:
\begin{equation}
  {\cal U}_Y = {\epsilon_0 \over 2}\int  \left[ \E^2  + 
\kappa^{-2} {( \div \E - \rho/\epsilon_0) ^2 } \right] \dV \label{noconst}\,.
\end{equation}
The analytic treatment of this energy is simpler than the constrained
functions that we have considered until now; the functional is
Gaussian in the field $\E$.  Clearly the limit $\kappa\rightarrow 0$
leads to a constrained electric field.  Taking the functional
derivative of Eq.~(\ref{noconst}) with respect to the electric field
we find
\begin{equation}
\kappa^2 \E - \grad \div \E = - \grad \rho/\epsilon_0
\end{equation}
for the stationary points of the energy function.  This equation is
best treated by separating the transverse and longitudinal components
of $\E$.  The longitudinal component of the electric field gives rise
to the Yukawa potential at the minimum of Eq.~(\ref{noconst}).  The
transverse components of $\E$ do not couple to the density.

When $\kappa$ is small the screening length becomes large and the
field becomes more and more constrained by Gauss' law.  Freely chosen
Monte-Carlo moves with Eq.~(\ref{noconst}) then become inefficient
since small amplitude trials are needed. It becomes more efficient to
separate the Monte-Carlo moves into two classes, those which conserve
$\div \E -\rho/\epsilon_0$ and those which modify it.

\section{Off Lattice interpolation and Relation to Plasma simulations}
We now formulate the algorithm when the particles are moving in the
continuum but with charges interpolated to the cubic grid. We shall
see that the simplest generalization of our method is closely related
to numerical methods used in plasma simulation.

Consider a single particle within a cell of the grid. We here take
$\Mesh=1$ and consider a cube such that $0<\{x,y,z\}<1$.  The
simplest interpolation of the charges to the lattice is a linear
interpolation so that the charge at the site $(0,0,0)$ is equal to $e
(1-x) (1-y) (1-z)$. Interpolating to the eight sites of the enclosing
cube preserves both the total charge and the dipole moment.

When the particle moves the rate of change of the density at the
origin is
\begin{eqnarray}
F(t) = -e [ (1-x) (1-y) \dot z &+& (1-x) (1-z) \dot y \nonumber \\
  &+& (1-y) (1-z) \dot x  ]
\label{currents} \,.
\end{eqnarray}
This variation has a simple, direct interpretation as currents in the
links:
\begin{equation}
J_z =e (1-x) (1-y) \dot z
\label{jz}
\end{equation}
is the current in the $z$ direction on the link between $(0,0,0)$ and
$(0,0,1)$.
 
Eq.~(\ref{currents}) is a perfect differential so that when the
particle is transported around a loop
\begin{equation}
\int F(t) dt=0 \label{loop}\,,
\end{equation}
implying that the interpolated charge returns to its initial value;
this is a clearly needed consistency requirement for the current.

Consider now Monte-Carlo: We need to generalize the expression for
\begin{math}
  J_z
\end{math}
to finite displacements, while preserving Eq.~(\ref{loop}).  Take a
path that starts at $(x_i,y_i,z_i)$ and finishes at
$(x_{i+1},y_{i+1},z_{i+1})$ after a time $t=1$. We integrate the
expression, Eq.~(\ref{jz}) for the link current to find
\begin{eqnarray}
\Delta e_z/e  &=& \int_0^{1} (1- x_i - v_{x} t) 
 (1- y_i - v_{y} t) v_{z} \; {\rm d}t \nonumber \\
 &=&  (1-x_i)(1-y_i) v_{z}  
\nonumber \\ &-&  {(v_{x} (1-y_i) +   v_{y} 
(1-x_i) ) v_{z}  \over 2} \nonumber \\
&+&{ v_{x} v_{y} v_{z}  \over 3} \,,
\end{eqnarray}
where $v_x = x_{i+1} -x_i$.  By construction a series of such steps
satisfies an expression equivalent to Eq.~(\ref{loop}).  The
transfered charge $\Delta e_z$ leads to a corresponding modification
of the electric field on the link $E \rightarrow E - \Delta
e_z/\epsilon_0$.

While being a low order interpolation scheme Villasenor and
Buneman\cite{buneman} showed that such a Gauss' law conserving scheme
allows long time, stable integration of plasma equations.  Many other
(non-conserving) methods need a periodic correction step in which
Poisson's equation is solved to re-initialize the longitudinal
electric field, leading to complicated and slow code. We again see the
central role played by Gauss' law in the efficient simulation of
charged systems; codes based on Buneman's TRISTAN  would seem to
be the closest analogy in the literature to our algorithm.

For the simplest, low precision modeling of polymers and
polyelectrolytes we expect that this present scheme will already be
useful.  The scheme needs to be improved by interpolation over several
cells in order to be competitive in accuracy with those that are
currently used in the soft condensed matter simulation
community\cite{darden}.  We present one possible algorithm in an
accompanying paper\cite{joerg} on charge and current interpolation.

\section{Inhomogeneous Media}

The forces on charges in the presence of inhomogeneous dielectric
media are complicated; a charge is attracted towards regions of high
$\epsilon$. Such forces are well known to be important in the
structuring of charge distribution around proteins \cite{perutz} and
at surfaces.

The electric energy of an arbitrary inhomogeneous medium is given
\cite{landau} by
\begin{equation}
{\cal U} = \int d^3 \r {{\D}^2 \over 2 \epsilon(\r)}\,,
\label {dielec}
\end{equation}
where the electric displacement, $\D= \epsilon(\r) \E$, obeys the
constraint
\begin{math}
  \div \D = \rho \,.
\end{math}
The generalized Poisson equation found from this constrained
variational problem is
\begin{equation}
\div (\epsilon( \r) \grad \phi ) = -\rho \label{genpoisson}\,,
\end{equation}
so that the electrostatic energy is given by
\begin{equation}
{\cal U}_{elec} = \int d^3 \r {\epsilon(\r) \over 2} ({\grad \phi_p })^2\,. 
\label {dielec2}
\end{equation}
The solution to the constraint equation in non-periodic media is
\begin{equation}
\D = -\epsilon(\r) \grad \phi +\curl \Q \,.
\end{equation}
As for the the case of homogeneous media the energy Eq.~(\ref{dielec})
can be written as a sum of two terms corresponding to electrostatic
and fluctuation contributions. 

The statistical mechanical treatment of inhomogeneous dielectric media
is however more complicated than the case of homogeneous systems: We
have seen that when $\epsilon$ is a constant the contribution of the
transverse fluctuations to the free energy is independent of the
positions of the particles; the algorithm gives rise to pure Coulomb
interactions between particles.  Here we shall show that in addition
to the electrostatic interactions, Eq.~(\ref{dielec2}), we also
generate extra dipole-dipole interactions.

We shall show that the fluctuation potentials are expected on general
grounds in inhomogeneous media, they correspond to the Keesom
potential\cite{keesom} varying as $1/r^6$ between fluctuating
permanent dipoles; These interactions should be distinguished from the
van der Waals interaction (also in $1/r^6$) which has its origins in
quantum mechanics.  \footnote{We also note that unlike van der Waals
  interactions the Keesom potential in the presence of free charges is
  screened over a distance of one half of the Debye length.}

\subsection{Dipolar forces}

In this section we consider the fluctuation potentials in the absence
of charges, then:
\begin{equation}
{\cal Z} = \int {\cal D} \D\, 
e^{-\int {\beta \D^2\over 2 \epsilon(\r)} \dV}\,
 \prod_{\bf r} \delta(\div \D(\r) ) \label{dz} \,.
\end{equation}
This energy is a function of the position of the particles due to the
presence of $\epsilon(\r)$ in the denominator of the energy.

We will proceed by firstly re-writing the constrained field $\D$ in
terms of a projection operator. This is easily done in Fourier space
where one can trivially separate longitudinal and transverse
components of a vector field. We then perform an inverse Fourier
transform to find the analytic form of the projection operator in real
space. Rather surprisingly we find that this projection operator is
closely related to dipolar interactions which vary in $1/r^3$. An
expansion to second order in the fluctuations in the dielectric
constant then gives a contribution to the free energy which naturally
decreases as $1/r^6$.

One calculates a transverse field from a general unconstrained field
$\D$ by subtracting off the longitudinal component:
\begin{equation}
\D_t(\q) = \D(\q)-{{\q ( \q.\D(\q))} \over q^2}
\end{equation}
Since we extract the transverse component by multiplication in Fourier
space the same operation can be written as a convolution in real
space:
\begin{equation}
\D_t (\r) = \int d^3\r_1 \;  \O(\r,\r_1) \D(\r_1) 
\end{equation}
with a real space projection operator $\O$.  This inverse Fourier
transform is a little tricky to perform: The projection operator does
not decay ``nicely'' for large $q$.  We proceed more carefully using
various operator identities:

We consider the function $\v(\r) \cdot \D_t(\r)$ for an arbitrary
vector $\v(\r)$. We note that $i{\bf q}\cdot$, $i{\bf q}$ and $1/q^2$
are respectively the Fourier transforms of $\div$, $\grad$ and $1/4\pi
r$. This results in
\begin{equation}
\v(1) \cdot \D_t(1) = \int \v(1) \left (\delta(1,2) + {1\over 4 \pi r_{12 }}
 \grad \div \D(2) \right) \; d2   
\end{equation}
We have used a common shorthand notation replacing $\r_i$ by the index
$i$.  We now integrate by parts twice (using the fact that the
operators $\div$ and $-\grad$ are mutually adjoint) to transfer the
derivatives from $\D$ to ${\v}/r$ while discarding surface terms:
\begin{equation}
\v(1) \cdot \D_t(1)
= \int \D(2) \left ( \delta(1,2)+ \grad 
\div {1\over 4 \pi r_{12}}\right)\v(1)   d2   
\end{equation}
 The operators ``$\grad$'' and ``$\div$'' act on
the ``$2$'' degrees of freedom so that ${\bf v}(1)$ is to be
considered as a constant. One simplifies using the identity
\begin{eqnarray}
\grad ( \v \cdot \grad (1/r))
&=& (\v \cdot \grad) (\grad (1/r)) \\
&=&{3 ({\bf v}\cdot \hat \r )\hat \r - {\bf v} \over r^3} \quad {\rm for }\;{r\ne0\nonumber }
\end{eqnarray}
leading to
\begin{equation}
\O(\r)= {2\over 3}  \delta(\r) {\bf I}  + 
{3\ket{\hat \r}\bra{\hat \r} - {\bf I} \over 4 \pi r^3} \,.
\end{equation}
${\bf I}$ is the unit matrix.  The coefficient of the delta function
is a little subtle but is found by noting that the trace of an
operator is independent of the basis: A projection operator which
selects the two transverse degrees of freedom has trace two.

The transverse projector is closely related to dipolar interaction
between particles. For instance the electrostatic interaction between
two dipoles ${\bf p}_1$ and ${\bf p_2}$ when $\r_1 \ne \r_2$ is just
\footnote{ One oddity is that $\O$ is the magnetic dipole operator not
  the electric dipole operator \cite{jackson}; they differ in the
  coefficient of the $\delta$-function.}
\begin{equation}
{\cal U}_{1,2} = -{\bf p}_1 \O_{1,2} {\bf p}_2/\epsilon_0
\end{equation}
as is discussed in many standard texts\cite{jackson}.  $\O$ has the
useful property
\begin{equation}
\int \O(0,1) \cdot \O(1,2)\; d^3 1 = \O(0,2)
\label{projection}
\end{equation}
since it is a projection operator and thus idempotent.  Here
``$\cdot$'' signifies matrix multiplication of the operator $\O$.

We now proceed by perturbation theory in fluctuations in the
dielectric constant. Let us consider two small inhomogeneities in the
dielectric properties separated by a large distance; the background
dielectric constant is $\epsilon_0$. One expands the partition
function to second order in the perturbation.  Each fluctuation in the
dielectric constant has associated with it an amplitude
$(1/\epsilon(\r) - 1/\epsilon_0) \approx -\delta
\epsilon/\epsilon_0^2$.

The second order perturbation in the partition function due to
interaction between two sites ``$a$'' and ``$b$'' is found to be
\begin{eqnarray}
\Delta_2(a,b)= {\beta^2 \over 4 \epsilon_0^4} 
 \int&&  \delta \epsilon(a) \delta \epsilon(b) \nonumber \; d^3\{1 2 3 4\}\\
 &\times&{\D} ( 1)\O(1,a) \cdot \O(a,2) \D(2)  \nonumber \\
  &\times& {\D} ( 3) \O(3,b) \cdot \O(b,4) \D(4)
\end{eqnarray}
which needs to be averaged with the statistical weight from
Eq.~(\ref{dz}). The interesting pairings are
\begin{math}
  \langle \D(1) \D(4)\rangle \langle \D(2) \D(3) \rangle
\end{math}
and
\begin{math}
  \langle \D(1) \D(3)\rangle \langle \D(2) \D(4) \rangle
\end{math}.
To perform the integration over transverse modes we introduce the
modified energy functional
\begin{eqnarray}
\bar {\cal U}&=& \int \bar {\D^2 \over 2 \epsilon_0} d^3 \r\\
\bar \D(\q) &=& \left(1 + \lambda \ket{\q} \bra{\q}\right ) \D(\q) \,. 
\end{eqnarray}
In the limit of $\lambda$ large and positive all longitudinal
fluctuations are suppressed in the measure, however we can now perform
usual, unconstrained perturbation theory with this Gaussian energy. In
the limit of large $\lambda$ the appropriately normalized correlators
become
\begin{equation}
\langle D_i(1) D_j(2) \rangle = \epsilon_0 \kt \O_{i,j} (1,2) \,.
\end{equation}

The calculation is much simplified by making use of the identity
Eq.~(\ref{projection}). The free energy cost of dielectric
inhomogeneity is given by
\begin{equation}
F = - \kt \ln {\cal Z} = F_0 +  V(a,b)\,,
\end{equation}
where $F_0$ is the free energy of a uniform dielectric medium. The
potential between particles is found to be
\begin{eqnarray}
V(a,b)(r) &=& -{\kt v_a \delta 
 \epsilon_a \;v_b \delta \epsilon_b\ \over 2 \epsilon_0^2}
 {\rm Tr}\, \O^2 (a,b)\\
           &=& -{ 3 \kt v_a\delta \epsilon_a 
\; v_b \delta \epsilon_b \over (4 \pi \epsilon_0)^2 r_{a,b}^6} \,,
\end{eqnarray}
with $v_i$ the volume of the dielectric inhomogeneity.  Substituting
that the relative dielectric constant for a weakly dipolar material
(with $\epsilon_i/\epsilon_0-1$ small) is given by
\begin{equation}
\epsilon_i/\epsilon_0 = 1+ {\p_i^2 n_i \over 3 \epsilon_0 \kt}  \,,
\end{equation}
with $n_i$ the density of dipoles of strength $\p_i$ we recognize the
Keesom potential \cite{keesom} between pairs of thermally agitated
permanent dipoles
\begin{equation}
V_K = - {\p_a^2 \p_b^2 \over 3 \kt (4 \pi \epsilon_0 r^3)^2}\,.
\end{equation}
Note the distinction between free energy and internal energy for the
Keesom potential. The internal energy is given by
\begin{equation}
u_K= {\partial \over \partial \beta} \beta F ,
\end{equation}
When the dipole moments are independent of the temperature $u_K=2V_K$.
If $\epsilon_i$ is independent of the temperature $u_K=0$; the
potential is purely entropic.

The presence of these dipole terms seems inevitable in the Monte-Carlo
algorithm: Removing them requires the evaluation of a complicated
multibody determinant. The algorithm is closer in spirit to
simulations which replace the dielectric medium by an ensemble of
fluctuating Langevin dipoles \cite{levitt} than the classic, zero
temperature, solution of the Poisson equation.

\subsection{Simulations}
Let us discretize the energy Eq.~(\ref{dielec}) by placing the particles
on the vertices of a cubic network. To each lattice point we also
associate a dielectric constant $\epsilon_i$. The energy associated
with a link ``$i,j$'' is then defined to be
\begin{equation}
{\cal U}_{i,j} = \Mesh^3 {D_{i,j}^2 \over 4}
\left ({1\over \epsilon_i} +{1\over \epsilon_j} \right) \,.
\end{equation}
The constraint $\Mesh^2 \sum_j D_{i,j}= e_i$ is imposed as an
initial condition on $\D$ rather than $\E$.  The electric displacement
$\D$ is updated in a similar manner to that described above for the
electric field.
The algorithm is simple compared with direct solution of the Poisson
Eq.~(\ref{genpoisson}); in particular we note that Fourier methods are
do not diagonalize the Poisson equation when $\epsilon$ is a function
of the position.

\begin{figure}[htb!]
  \includegraphics[scale=.45] {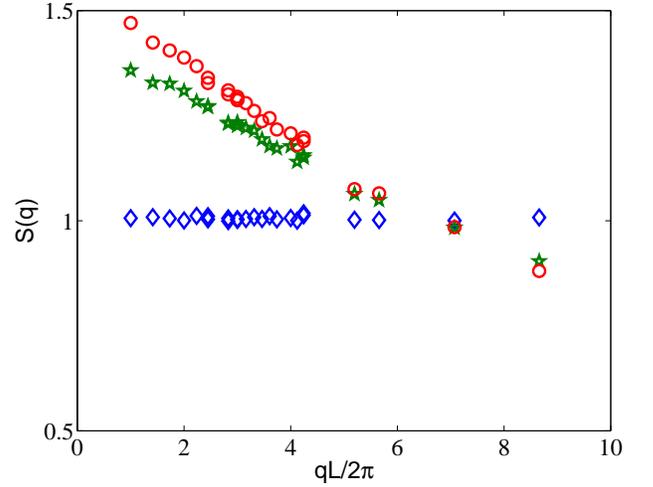}
\caption
{Plot of the density-density structure factor as a function of
  wave-vector.  1000 neutral particles simulated on a
  $15\times15\times15$ grid.  $\diamond$: $\epsilon_p=\epsilon_0$.
  $\star$: $\epsilon_p=5\epsilon_0$.  $\circ$:
  $\epsilon_p=0.2\epsilon_0$. For small $q$ the structure factor
  increases, reflecting attractive interactions between the particles
  when $\epsilon_p \ne\epsilon_0$. }
\label{tripple}
\end{figure}

We implemented the algorithm for a neutral lattice gas of dielectric
constant $\epsilon_p$ embedded in a background $\epsilon_0$ in the
absence of free charges. The density-density correlation functions
$S(q)$ were measured to characterize the effect of interactions,
 Fig.~(\ref{tripple}).
  As
expected when $\epsilon_p= \epsilon_0$ no structure appears in the
simulation.  For both $\epsilon_p< \epsilon_0$ and $\epsilon_p>
\epsilon_0$ the structure factor increases for small $q$. This can be
interpreted as being due to an attractive interaction between the
particles leading to clumpiness in the structure. 

\subsection{Low symmetry materials}

In low symmetry materials the dielectric constant is a tensor, with up
to three distinct eigenvalues and eigenvectors. The
algorithm generalizes to this case;
in such materials the constraint equation is
\begin{equation}
\div \D = \rho\,,
\end{equation}
where $ \D = \epsilon \E$; $\epsilon$ is a matrix to that $\D$ and
$\E$ need no-longer be parallel.  The electric energy 
\begin{equation}
{\cal U} = {1\over 2} \int \D \epsilon^{-1} \D \dV \,.
\end{equation}
$\epsilon^{-1}$ denotes the matrix inverse.

One possible application of this discretization is in liquid crystals
where the order parameter manifests itself as a dielectric anisotropy.
Models are  easily constructed within a mean field
description. This permits numerical study of transitions driven
by electric fields.

\section{Conclusions}

This paper has presented local algorithms for the simulation of
charged systems.  Our methods are strongly inspired by the observation
that Maxwell's equations allow a local evolution of field degrees of
freedom leading to the Coulomb interaction.  It introduced dynamics
which conserve $\div \E -\rho/\epsilon$ while starting the simulation
with this constraint satisfied.  The Monte-Carlo algorithm eliminates
the magnetic degrees of freedom from Maxwell's equations leading to
pure electrostatic interactions.

The method is rather general and can be used in many situations: for
instance application of the constrained formalism to path integral
Monte-Carlo should be equally straight forward as the cases treated
here. The main weaknesses that we see in the algorithm are
\begin{itemize}
\item Incompatibility of constraint preservation with simulations in
  the grand-canonical ensemble, so useful in the study of phase
  behavior.
\item Low efficiency with large strongly charged objects undergoing
  coherent motion, either due to rigidity or the use of smart global
  moves.
\end{itemize}

The code used in the simulations is available from the author.

\vskip 0.5cm I would like to thank Ralf Everaers for many discussions,
essential in the formulation of these algorithms.  \bibliography{mc}
\end{document}